\documentclass[%
 aip,
 amsmath,amssymb,
 reprint,%
]{revtex4-1}

\usepackage{graphicx}
\usepackage{dcolumn}
\usepackage{bm}

\usepackage[utf8]{inputenc}
\usepackage[T1]{fontenc}
\usepackage{mathptmx}

\usepackage{etoolbox}
\usepackage{soul}

\makeatletter
\def\@email#1#2{%
 \endgroup
 \patchcmd{\titleblock@produce}
  {\frontmatter@RRAPformat}
  {\frontmatter@RRAPformat{\produce@RRAP{*#1\href{mailto:#2}{#2}}}\frontmatter@RRAPformat}
  {}{}
}%
\makeatother
\begin{document}

\preprint{AIP/123-QED}

\title[Hydrogen Passivation Effects on Spatially Resolved Charge Trap Densities in Si(100)-SiO$_2$]{Hydrogen Passivation Effects on Spatially Resolved Charge Trap Densities in Si(100)-SiO$_2$}
\author{Adam J. Czarnecki}
\email{Adamc@physics.mcgill.ca.}
\altaffiliation[]{These authors have contributed equally to this work.}
\affiliation{Department of Physics, McGill University, Montreal, Quebec, Canada}%

\author{Nikola L. Kolev}%
\email{Nikola.Kolev.21@ucl.ac.uk.}
\altaffiliation[]{These authors have contributed equally to this work.}
\affiliation{London Centre for Nanotechnology, University College London, London, United Kingdom}%
\affiliation{Department of Electronic and Electrical Engineering,
University College London, London, United Kingdom}

\author{Patrick See}
\affiliation{National Physical Laboratory, Teddington, United Kingdom}

\author{Nick Sullivan}
\affiliation{Department of Physics, McGill University, Montreal, Quebec, Canada}%

\author{Wyatt A. Behn}
\affiliation{Department of Physics, McGill University, Montreal, Quebec, Canada}%

\author{Neil J. Curson}
\affiliation{London Centre for Nanotechnology, University College London, London, United Kingdom}%
\affiliation{Department of Electronic and Electrical Engineering,
University College London, London, United Kingdom}%

\author{Taylor J.Z. Stock}
\affiliation{London Centre for Nanotechnology, University College London, London, United Kingdom}%
\affiliation{Department of Electronic and Electrical Engineering,
University College London, London, United Kingdom}%

\author{Peter Grütter}
\affiliation{Department of Physics, McGill University, Montreal, Quebec, Canada}%

\date{28 May 2025}

\begin{abstract}
As silicon-based devices continue to shrink to the nanoscale, traps at the Si-SiO$_2$ interface pose increasing challenges to device performance. These traps reduce channel carrier mobility and shift threshold voltages in integrated circuits, and introduce charge noise in quantum systems, reducing their coherence times. Knowledge of the precise location of such traps aids in understanding their influence on device performance. In this work, we demonstrate that frequency-modulated atomic force microscopy (fm-AFM) allows the detection of individual traps. We use this to study how sample preparation, specifically the introduction of a buried hydrogen termination layer, and post-processing annealing in forming gas (N$_2$+H$_2$), affects the density of donor-like traps in Si(100)-SiO$_2$ systems. We spatially map and quantify traps in both conventionally prepared (“pristine”) silicon samples and those processed under ultra-high vacuum for hydrogen resist lithography (HRL). We confirm previous studies demonstrating hydrogen passivation of traps and find that hydrogen termination further reduces the donor-like trap density. We also observe a significant reduction in two-level donor-like traps in the hydrogen-terminated samples compared to pristine silicon samples. These findings suggest that HRL-prepared silicon may offer advantages for high-performance nanoscale and atomic-scale devices due to reduced trap densities.
\end{abstract}

\maketitle

\section{\label{sec:level1}Introduction}

Silicon remains the foundational material for modern electronics, from everyday computing to emerging quantum technologies. As devices shrink to the nanoscale, their performance becomes increasingly sensitive to material imperfections, particularly at the Si-SiO$_2$ interface. These defects include interfacial traps, arising from the mismatch between the silicon and oxide crystals, as well as oxide traps within the oxide layer itself. Such traps can disrupt electronic properties by introducing charge noise, lowering carrier mobility, shifting threshold voltages, and increasing leakage currents \cite{PbCenterKato, threshVoltageShiftCampbell, defectNoiseKirton, CowiePNAS}. These issues are particularly concerning for field-effect devices such as metal–oxide–semiconductor field-effect transistors (MOSFETs) \cite{nanoMosfetChen}, as well as emerging silicon-based quantum technologies \cite{qubitCoherenceWang, PCoherenceAmbal}.

Reducing trap density is therefore a key challenge in the fabrication of high-performance devices. Hydrogen and deuterium passivation of interfacial traps in Si-SiO$_2$ systems via thermal annealing has been well studied \cite{StesmansSi100Passivation, StesmansSi111Passivation, Brower, DBPassivationCartier, HydrogenReactionsPantelides, deuteriumAnnealLyding}. The experimental methods typically used to characterize trap densities in these studies include electron spin resonance and nuclear magnetic resonance. These techniques quantify average trap densities over large sample areas. As device dimensions continue to shrink, the spacing between individual traps can become comparable to, or even larger than, the device's critical dimension. Consequently, it is necessary to know the precise location of traps to understand their influence on device performance, as global trap density is less relevant than the exact position with respect to a device.

In this work, we demonstrate that frequency-modulated atomic force microscopy (fm-AFM) allows the detection of individual traps. We use this to study how sample preparation—specifically hydrogen termination and post-processing annealing in forming gas (N$_2$+H$_2$)—affects the density of donor-like traps in Si(100)-SiO$_2$ systems. 
We spatially map and quantify traps in two types of silicon surfaces: conventionally prepared (“pristine”) silicon and hydrogen-terminated Si(100) prepared under ultra-high vacuum conditions, each with a native oxide layer. The latter is commonly used as a substrate in hydrogen resist lithography (HRL), a scanning tunneling microscopy-based technique for atomic-scale patterning \cite{Lyding}. HRL has shown promise for building single-atom transistors, quantum dots, and other atomically precise devices \cite{SingleAtomTransistorSimmons, atomQuantumDevicesWyrick, siliconLogicHuff, atomicMemoryAchal, quantumDotDeviceSimmons, quantumSimDeviceSilver}. However, the impact of hydrogen termination on trap densities, in these atomic-scale devices, has not been widely studied. This work examines how hydrogen termination and post-processing annealing influence trap densities at the Si(100)-SiO$_2$ interface. Our results highlight how fabrication and processing choices affect interface quality, offering insights for improving both conventional and quantum device architectures. We also demonstrate the utility of fm-AFM as a non-destructive tool for characterizing interfacial defects at the nanoscale.

\section{Experimental Methods}

All experiments were performed on silicon samples taken from the same initial wafer, which was boron-doped $\left(9.15\times10^{14}/\text{cm}^{3}\right)$ and 500 µm thick. The preparation of clean Si(100) surfaces, and subsequent hydrogen termination, took place in an ultra-high vacuum (UHV) chamber, with a base pressure of around $4 \times 10^{-10}$ mbar. A clean Si(100) surface was prepared by direct current heating to 600 $^{\circ}$C for 8 hours, followed by multiple flash anneals at 1200 $^{\circ}$C, and then slow cooling over 30 minutes to allow the surface to relax into a Si(100)-$2\times 1$ surface reconstruction with minimal defects. H-termination was performed by exposing the surface to an atomic hydrogen beam while using direct current heating to maintain a sample temperature of 320 $^{\circ}$C. The beam was generated by thermally cracking H$_2$, at a H$_2$ partial pressure of $5\times 10^{-7}$ mbar. Molecular beam epitaxy was then used to cap the sample in 3 nm of epitaxial silicon. The first 10 monolayers (the locking layer) was grown at room temperature, followed by a rapid thermal anneal at 470 $^{\circ}$C for 15 s. The rest of the silicon was then grown while the sample temperature was kept at 250 $^{\circ}$C. This is a common method for encapsulating devices fabricated using HRL, as it minimizes the diffusion of atomically precise positioned dopant atoms. The sample was subsequently removed from UHV to allow for a native oxide, of around 1 nm, to form at the surface, and diced into three parts.

To examine how annealing in a forming gas affects the density of Si(100)-SiO$_2$ traps, three types of post-fabrication anneals were studied: no anneal (control), N$_2$ gas, and N$_2$+H$_2$ forming gas. This was repeated for two types of Si(100) samples: unprocessed Si(100) - which was not cleaned in UHV or subsequently hydrogen terminated - and is referred to as “pristine silicon”, and Si(100) which was processed as one would for HRL \cite{Lyding, Stock:2020aa}, referred to as “H-terminated silicon”. The post-fabrication anneals were all performed 250 $^{\circ}$C for 5 minutes at atmospheric pressure, repeated four times for each sample. The ratio of the N$_2$+H$_2$ forming gas was 95\%/5\%.

\begin{figure}
\includegraphics[scale=0.5]{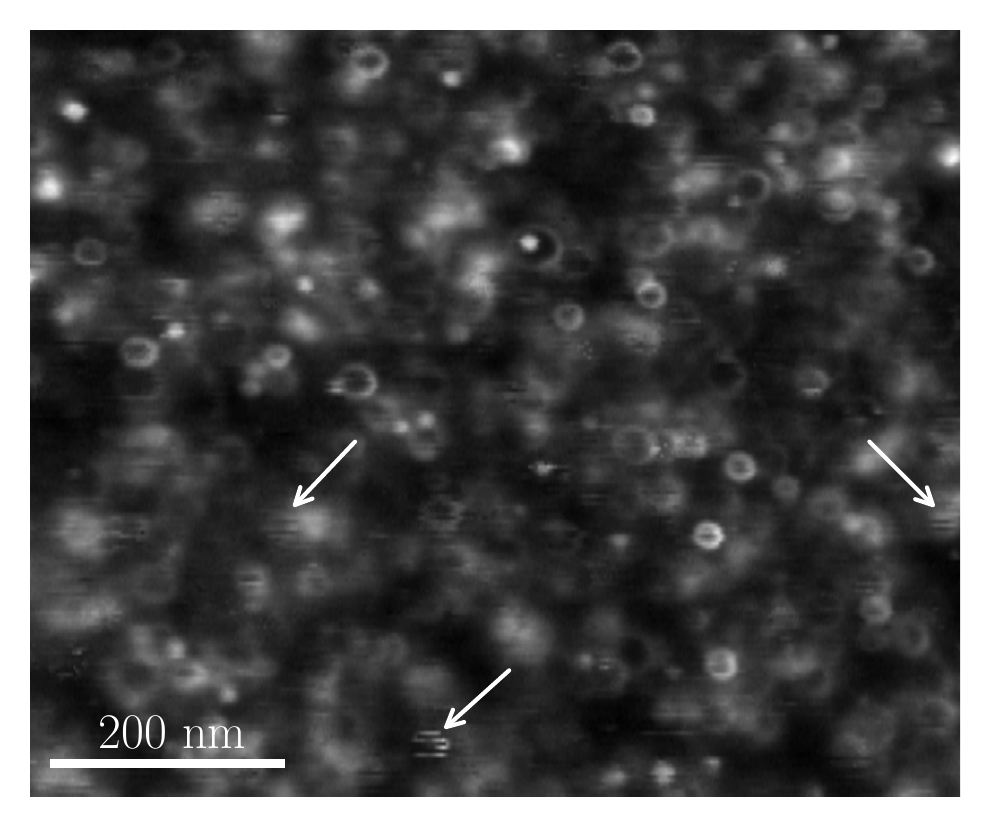}
\caption{\label{fig:black-white-pristine-scan} fm-AFM dissipation scans reveal rings due to traps. Shown is a dissipation scan done on the non-annealed pristine silicon sample, conducted at a bias of -3 V and $\Delta f$ setpoint of -300 Hz. The arrows point to three examples of traps showing two-level fluctuations. The b/w scale is 0:239 meV/cycle.}
\end{figure}

The spatial density of traps in the samples was characterized with fm-AFM using a metal-coated tip. Traps can be located by mapping the fm-AFM dissipation $F_{d}$ \cite{CowiePRL}. Areas of larger $F_{d}$ due to traps will manifest themselves as rings in a dissipation scan, with the rings centered at the trap locations. The rings are a result of the AFM response due to the ionization of a single trap \cite{CowiePRL}. An example of such a dissipation scan is shown in Fig.~\ref{fig:black-white-pristine-scan}, which was obtained from the non-annealed pristine silicon sample.

The ring size in dissipation scans depends on the tip bias and tip height \cite{CowiePRL, CockinsPNAS}. The samples studied here all have the same bulk doping, so the dominant electrostatic forces are expected to be constant across all samples. Thus, scanning the samples with the same frequency shift ($\Delta f$) setpoint is assumed to result in the same tip heights for all measurements and yield a consistent distribution of ring sizes in dissipation scans. While asymmetries in the ring shapes are likely due to the probe shape, slight variations in ring sizes may occur due to variations in trap depth and energy level.

Here, the tip was initially approached at zero bias and -10 Hz frequency shift setpoint. The $z$ feedback was then turned off, the bias adjusted to the desired value, and the $\Delta f$ setpoint adjusted to maintain the same $z$ position once the $z$ feedback was turned back on again. The bias and $\Delta f$ setpoint values were chosen so that rings were clearly visible in the dissipation scans, and kept constant across all samples to maintain a consistent ring size distribution. Negative tip bias allows for the imaging of donor traps, while positive bias allows for the imaging of acceptor traps\cite{CowiePRL}. A fraction of traps exhibited two-level fluctuations, as indicated by the arrows in Fig.~\ref{fig:black-white-pristine-scan}, and are discussed in further detail below.

\section{ Trap Density}

To assess the density of rings (and thus traps) in the dissipation scans, various computer vision methods were considered. Such techniques have been successfully applied to detect defects, molecules, or atoms in microscopy images \cite{zhu2024autonomous, shen2021multi}. For instance, Zhang et al. employed a combination of a YOLOv5 network and a ResNet-34 to locate and classify AFM imaging defects, achieving accuracies between 75\% and 95\% \cite{zhang2024AFM}. Based on preliminary evaluations and the success of these methods in similar contexts, we identified two approaches as the most effective - Otsu's pixel thresholding method to separate the ring regions from the background \cite{Otsu}, and a deep learning (DL) approach to locate and count the traps. Otsu's method automatically determines the pixel value threshold needed to separate the foreground from the background and does not require user-defined parameters. This results in consistent processing of all images. Measuring the change in the areas of the ring regions across the samples allows us to gauge how the trap density changes. However, it should be noted that the comparison of ring areas using Otsu's thresholding across samples assumes a consistent ring size distribution across all samples, regardless of the processing method. This assumption may not hold if hydrogen passivation preferentially affects certain types of donor-like traps (e.g., oxide or interfacial traps), potentially altering the ring size distribution in dissipation scans. To address this, we trained a YOLOv8 DL object detection model \cite{yolov8} on a synthetic dataset, to predict bounding boxes to identify and localize traps. This approach also provides the count and spatial distribution of traps. Full details on the synthetic dataset, training procedure, and model predictions are provided in the supplementary material. While this method offers some advantages over thresholding—particularly in spatial resolution—it is not without limitations: yielding some false positives and false negatives, and requiring far more computational resources. We present results using both of these methods, and discuss common trends as well as discrepancies. 

\begin{figure}
\includegraphics[scale=0.41]{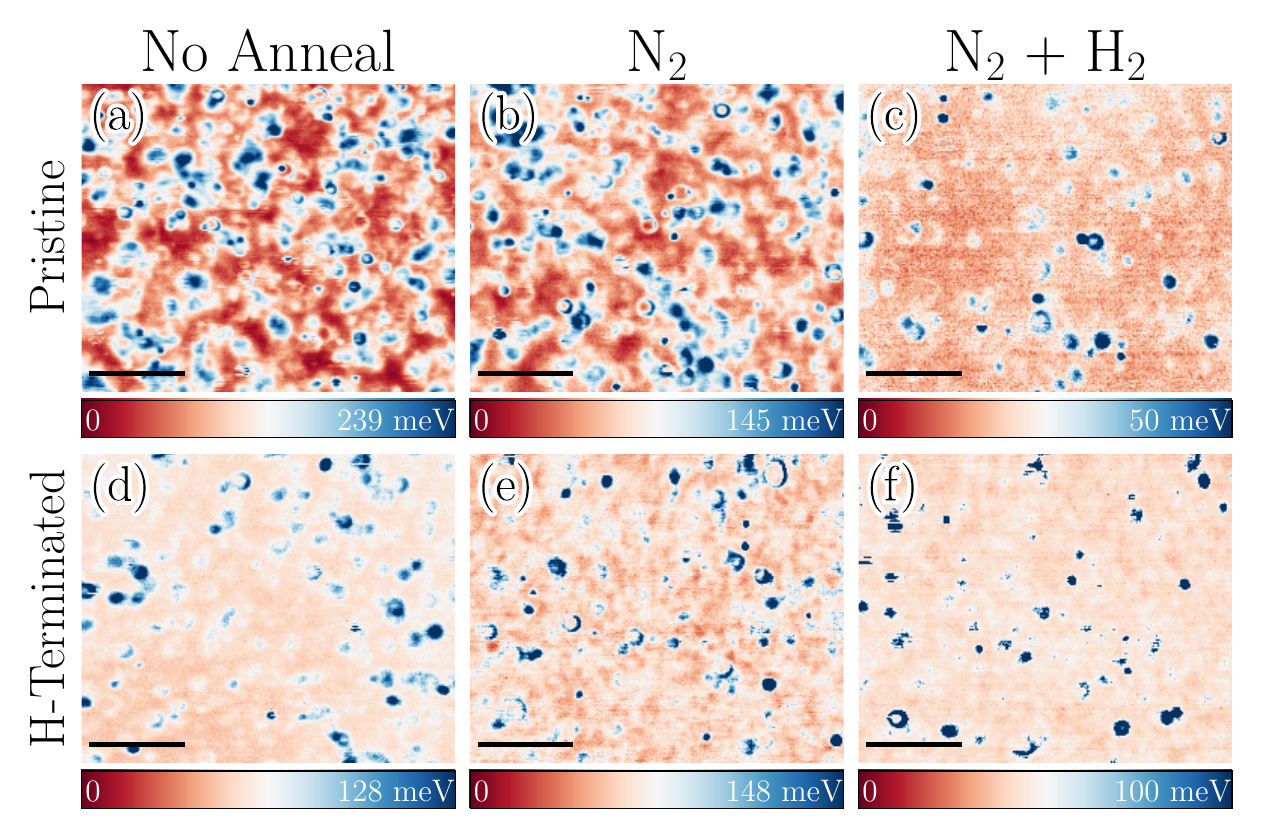}
\caption{\label{fig:pristine-scans} Negative bias dissipation scans reveal rings (blue regions) due to donor traps. Scans were conducted at a bias of -3 V and $\Delta f$ setpoint of -300 Hz. Scale bar is 200 nm. The fast scan axis is horizontal and slow scan axis is vertical. Horizontal scan speed was $\sim$20 nm/s. The scans in (a-c) were performed on the pristine silicon samples, while the scans in (d-f) were performed on the H-terminated silicon samples.}
\end{figure}

We performed dissipation scans on the samples at both positive and negative tip biases. We observe variations in density of donor traps (Fig.~\ref{fig:pristine-scans}) among the samples, but no variation in acceptor traps. The ring area fraction in dissipation scans performed with positive bias (i.e. acceptor traps) was found to be 0.25$\pm$0.01 across all samples. This work primarily discusses donor-like traps, and will simply be referred to as traps.

\begin{figure}
\includegraphics[scale=0.41]{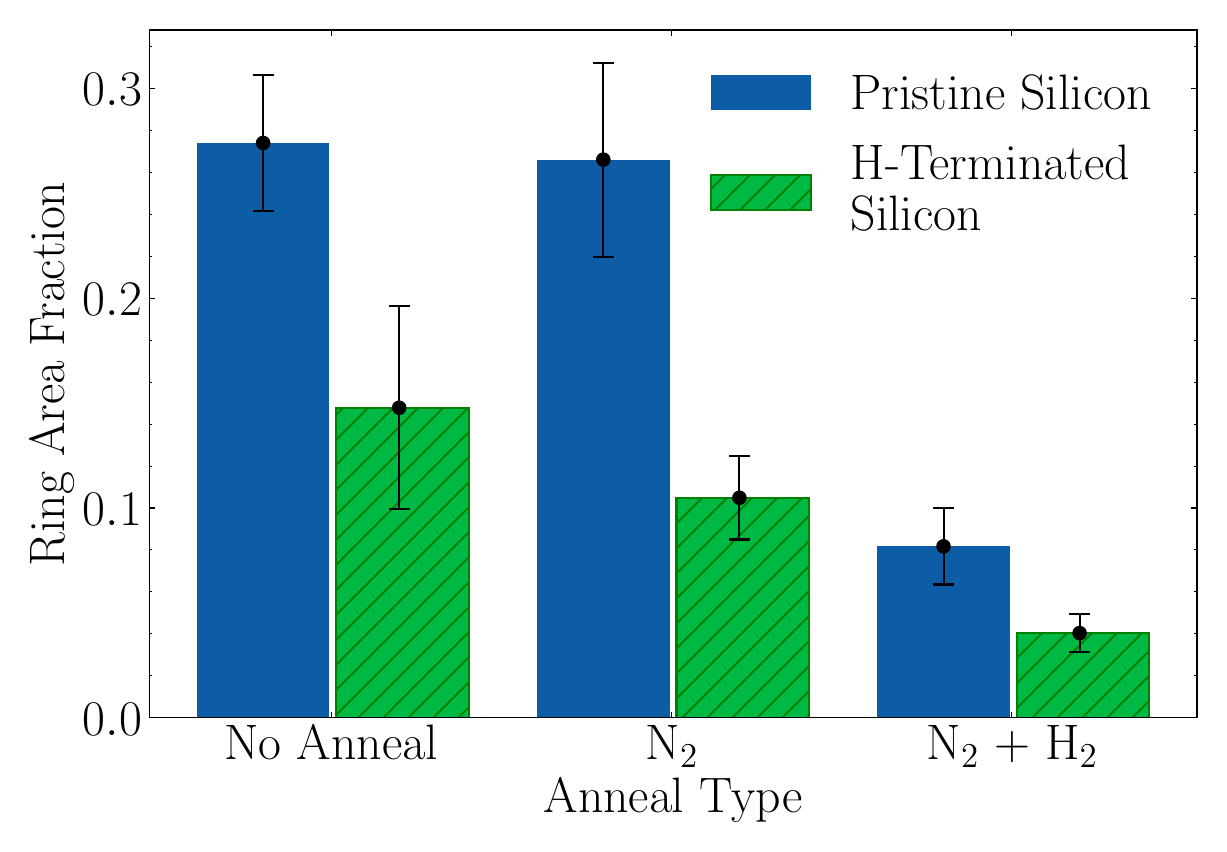}
\caption{\label{fig:ring-hist} Binarizing the scans shown in Fig.~\ref{fig:pristine-scans} reveals the areas covered by the rings. The uncertainties were calculated via a Monte Carlo windowing method, where quarter frames of the scans were randomly selected and the standard deviation of the measured ring areas was determined. }
\end{figure}

\begin{figure}
\includegraphics[scale=0.41]{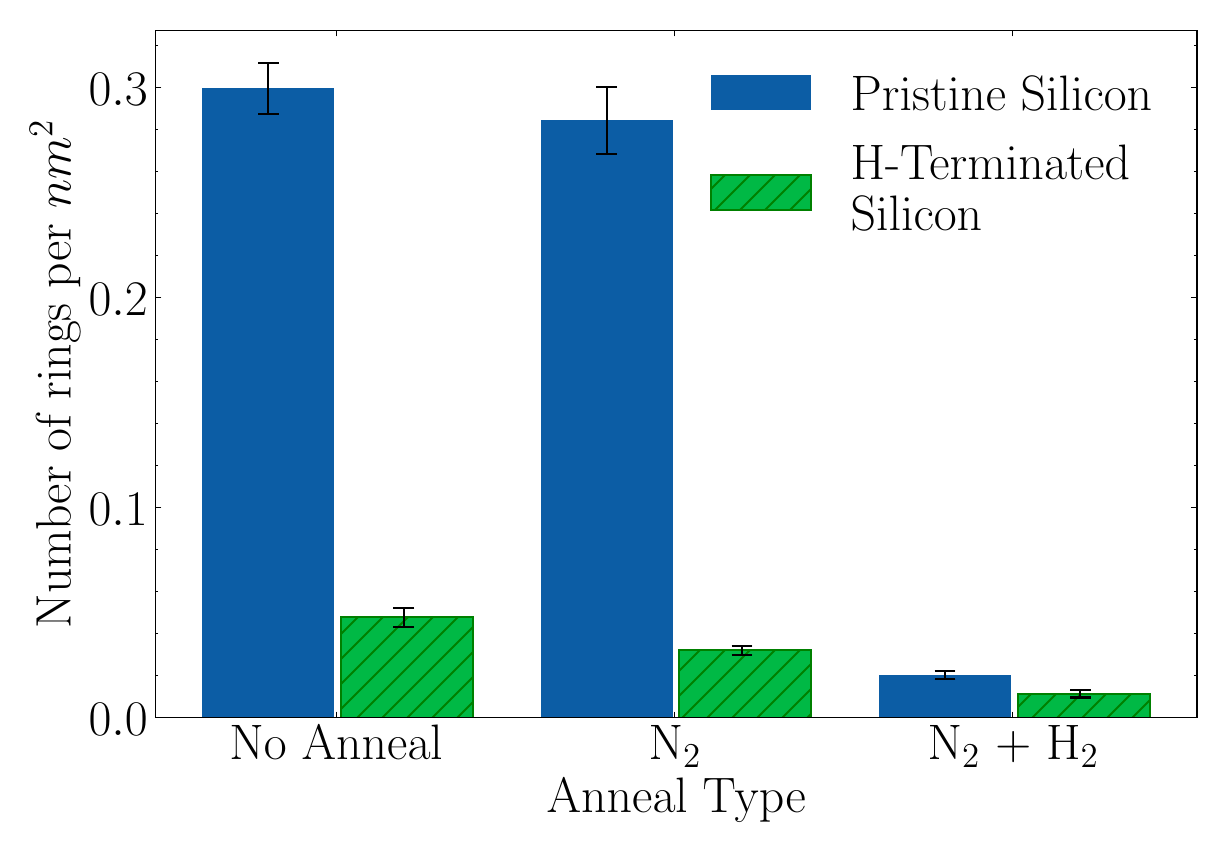}
\caption{\label{fig:ring-hist-ML} A YOLOv8 medium sized model was trained to predict bounding boxes for rings in the dissipation scans shown in Fig.~\ref{fig:pristine-scans}. The histogram shows the density of bounding boxes predicted by this deep learning model. Uncertainties were calculated by integrating dropout layers into the model and using the method presented by Gal et al. \cite{gal2016dropout}.}
\end{figure}

The dissipation scans performed in this work were over areas large enough (> 600 nm $\times$ 600 nm) to draw conclusions about the average trap density. Fig. ~\ref{fig:ring-hist} shows the results from Otsu's threshold analysis. To estimate the variation in the ring density calculated using Otsu's threshold, we measured the variance in ring density in 1000 randomly placed quarter sections of the scans, with wrapped boundary conditions. Given the high density of rings, we deemed these quarter sections to be representative of the overall coverage. Fig.~\ref{fig:ring-hist-ML} shows the results using the YOLOv8 DL method. By adding some randomness (using dropout layers during inference) to the YOLOv8 network it can be approximated as Bayesian approximation for deep Gaussian processes \cite{gal2016dropout}. This allows us to use Gaussian statistics to estimate the uncertainties in the network's predictions.

In general, we observe a greater reduction in traps when there is more hydrogen present in the sample processing, consistent with previous studies that have demonstrated the H-passivation of Si-SiO$_{2}$ traps \cite{StesmansSi100Passivation, StesmansSi111Passivation, Brower, DBPassivationCartier, HydrogenReactionsPantelides}. We observe that in both the pristine silicon samples and the H-terminated samples, annealing in N$_{2}$+H$_{2}$ forming gas leads to a reduction in the density of traps compared to the non-annealed and the N$_{2}$-annealed samples. Additionally, we observe that the H-terminated samples have a significantly lower trap density than the pristine silicon samples.  There is no statistical difference between the non-annealed and the N$_{2}$-annealed samples.

There is an open question as to what happens to the hydrogen layer when the silicon capping layer is deposited after hydrogen lithography in the H-terminated samples \cite{SiEpitaxyDeng}. We observe that the H-terminated samples have the lowest trap density. The overall reduction in trap density in the H-terminated samples relative to the pristine silicon samples suggests that the additional hydrogen helps further passivate donor traps.

\begin{figure}
\includegraphics[scale=0.41]{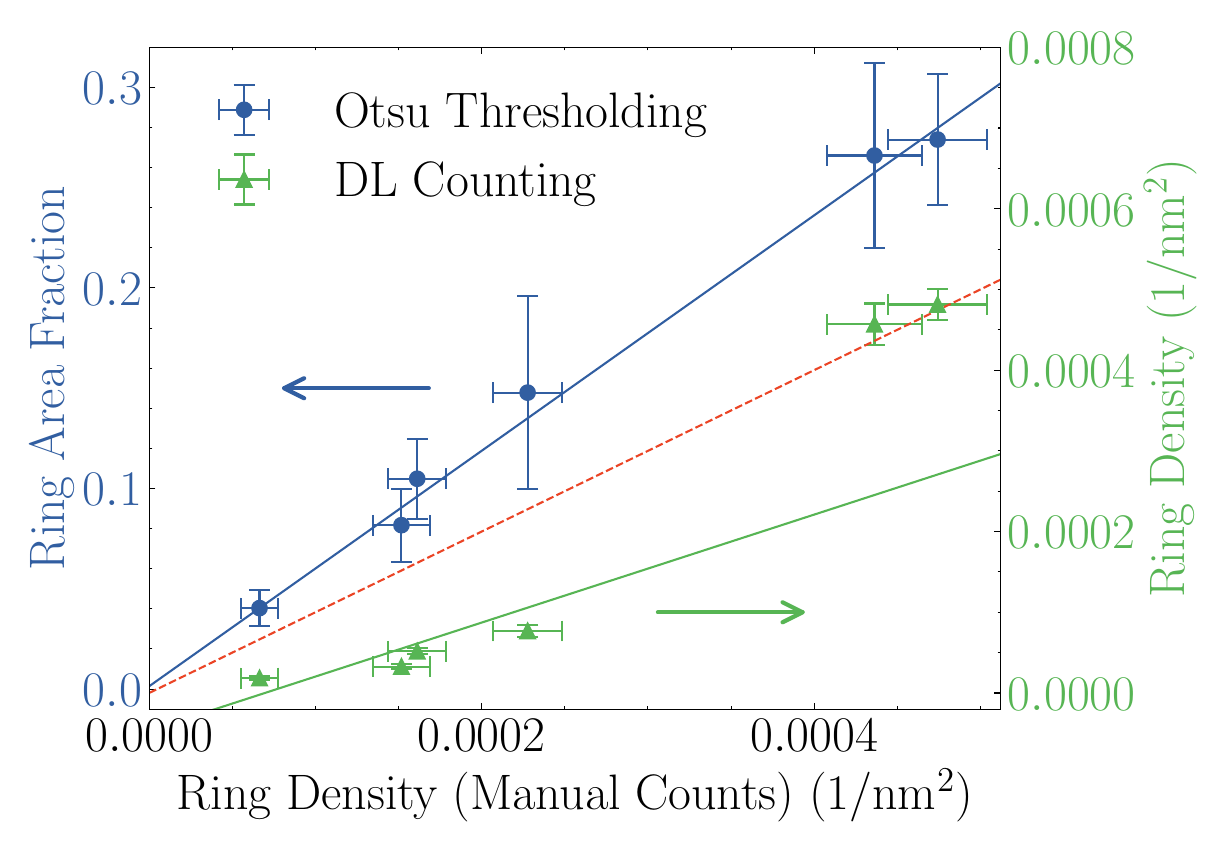}
\caption{\label{fig:manual_count_plot} Comparison of the Otsu thresholding and the DL methods of quantifying rings with manual counting. Uncertainties in the manual counting were assigned by assuming a Poissonian distribution of rings. Each counting method was fit by a linear function (solid lines). The Otsu thresholding method has a consistent linear trend, with a fit slope of 590$\pm$30 nm$^2$. The DL method accurately detects the number of rings in the scans with a large number of rings, while under counting in the scans with few rings. This is illustrated by the red dashed line, which has a slope of 1. The high density counts are in good agreement with this line, while the low density counts fall short of it.}
\end{figure}

To further validate the trends identified using the two methods of image analysis, we performed manual ring counting on the scans. A comparison of this manual counting to the two methods is shown in Fig.~\ref{fig:manual_count_plot}. Otsu's method follows a linear trend when compared to manual counting, with a fitted slope of 590$\pm$30 nm$^2$, offering a good conversion between ring area fraction and number of rings. In the case of DL vs manual counting, a line of slope 1 would indicate perfect agreement. We find the number of rings detected by the DL method is in good agreement with the manual counting in scans with a large number of rings, particularly the pristine silicon samples with no anneal and the N$_2$ anneal. However, the DL method detects a large number of false negatives (i.e., some rings are not detected), particularly in the dissipation scans of the H-terminated samples. Thus, the DL method underestimates the trap density in the H-terminated samples. This explains the discrepancy in the number of rings detected between the Otsu thresholding and DL methods. Further refinement of synthetic training data used in the DL method could help reduce false negatives.

Despite these discrepancies, all approaches to assessing trap ring density consistently showed that more hydrogen leads to greater trap passivation. Further details about the DL method, Otsu thresholding, and manual counting are provided in the supplementary material.

\section{Two-Level Trap Density} {\label{sec:two-level-traps}}

\begin{figure}
\includegraphics[scale=0.43]{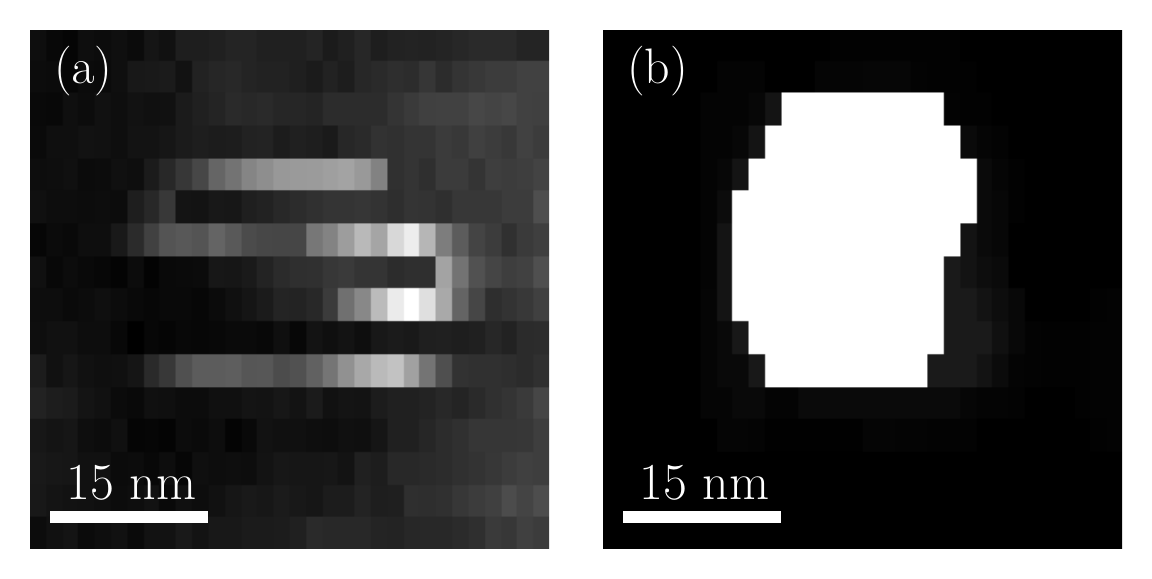}
\caption{\label{fig:two-level-ring} Taking the second derivative along the slow scan axis (here the y axis) reveals the locations of two-level traps. \textbf{(a)} An example of a two-level trap visible in Fig.~\ref{fig:pristine-scans}(a). \textbf{(b)} The second derivative with respect to the slow scan axis of the image in (a). A median filter was applied to join the pixels into one connected feature, necessary for quantification of features. The result was thresholded to make the region more prominent. Horizontal scan rate was $\sim$20 nm/s.}
\end{figure}

Some rings present in the dissipation scans correspond to defects that fluctuate between two levels, referred to as two-level traps \cite{CowiePNAS}. In addition to studying the density of all traps, we also examined how sample processing influences the density of these two-level traps. As the AFM tip moves across such a two-level trap, the trap may switch states, resulting in a change in the measured dissipation. This results in the corresponding dissipation ring to have a striped appearance, with stripes oriented along the fast scan axis. Some of these are highlighted in Fig.~\ref{fig:black-white-pristine-scan}, and a close-up example is shown in Fig.~\ref{fig:two-level-ring}.

\begin{figure}
\includegraphics[scale=0.41]{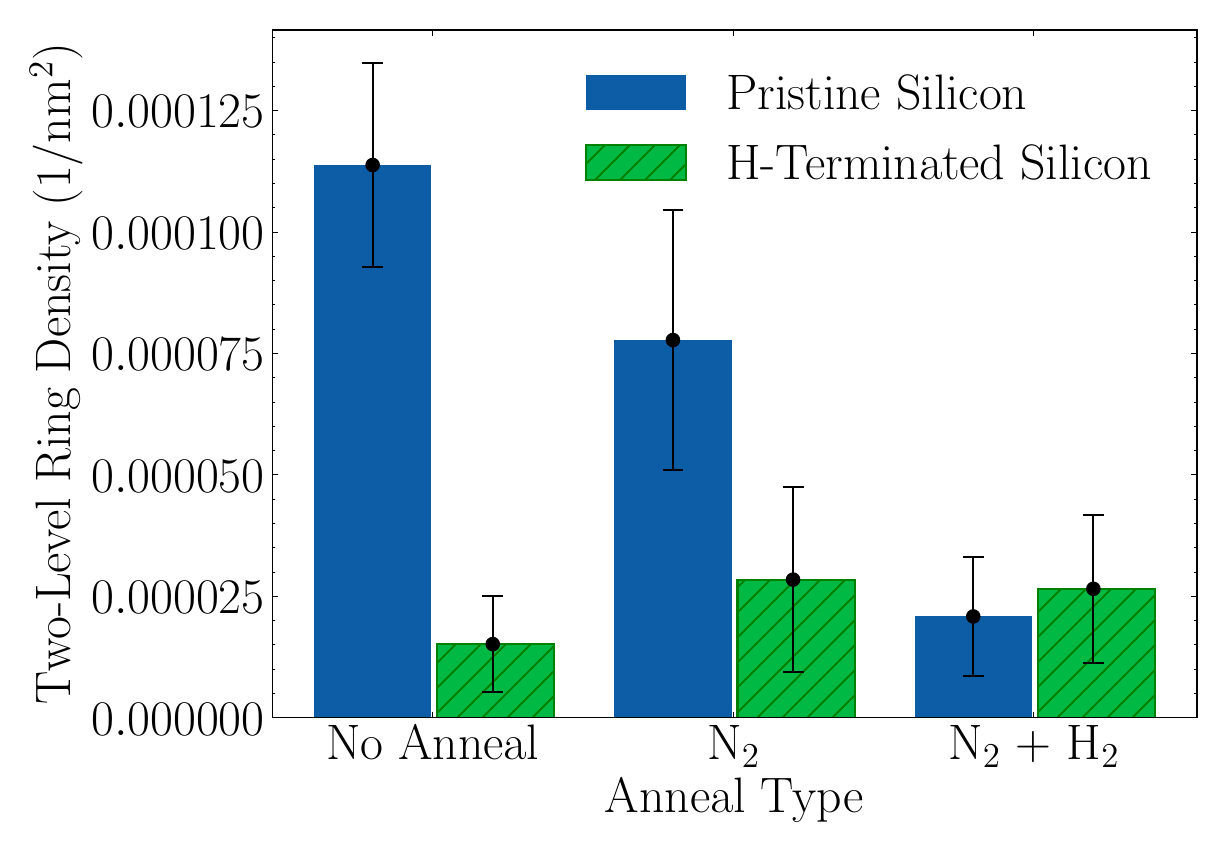}
\caption{\label{fig:two-level-hist} The histogram shows the spatial density of two-level traps across the samples. Given that very little overlapping or clustering of two-level traps was observed in the second derivative images, quantification was performed by counting disconnected features in the images, rather than adopting the method of measuring the areas of the features, used for Fig.~\ref{fig:pristine-scans}.}
\end{figure}

To assess the two-level trap density, we analyzed the second derivative with respect to the slow scan axis of the dissipation scans. Striped rings that correspond to two-level traps will appear as regions with a large second derivative. We used a similar pixel thresholding procedure as mentioned above to gauge how the density of two-level traps changes with sample processing. A quantification of the two-level traps is illustrated in Fig.~\ref{fig:two-level-hist}. We observe that in the pristine silicon samples, annealing in N$_{2}$ and N$_{2}$+H$_{2}$ forming gases reduces the number of two-level traps. As for the H-terminated silicon samples, we observe a very small number of two-level traps across all three samples. These two-level traps can be passivated by hydrogen, much like the static traps previously discussed \cite{CowiePNAS}.

It should be noted, that two-level traps occur with varying switching rates, and not all two-level traps present in the samples are detected in the dissipation scans presented here. In particular, two-level traps that switch much faster than the dwell time of the AFM tip at each pixel, as well as those that switch much slower than the time required to scan an entire trap, will not be distinguishable from static traps in the dissipation scans. By using variable scan rates it may therefore be possible to detect the presence of two-level traps at a wide range of frequencies. In this work however, the scan rate was kept constant, so the densities reported here are of two-level traps with switching rates within a consistent range, allowing for conclusions about the density trends.

\section{Conclusion and Outlook}

In summary, we investigated how the density of traps in the Si(100)-SiO$_{2}$ interface is processing, in particular, by exposure to hydrogen. This was done by performing fm-AFM dissipation scans on silicon samples processed in various forming gases, as well as H-terminated samples that had a layer of hydrogen below an epitaxially grown silicon layer. Using such AFM methodology allows for nanoscale spatial imaging of traps. We verified that an increased presence of hydrogen in the sample processing leads to greater passivation of donor-like traps. This is especially evident in H-terminated Si(100)-$2\times1$ samples that are used for fabricating devices via hydrogen resist lithography. We found that such H-terminated samples have significantly fewer traps than the pristine silicon samples, providing an answer to the open question of what happens to the hydrogen layer once a silicon capping layer is deposited on top. We conclude that this hydrogen contributes to the passivation of donor-like traps. We did not observe changes in the acceptor-like trap density; we therefore conclude that acceptor-like traps are not passivated by hydrogen. We also examined the occurrence of two-level traps, which are known to contribute to random telegraph noise in electronic and quantum devices \cite{CowiePNAS, volatileDefectsHydrogenShluger}, as well as loss in resonators and qubits \cite{TLSaSiHDefrance}. We observed that these two-level systems were strongly suppressed in hydrogen-rich environments, with H-terminated samples showing an 80\% reduction compared to pristine silicon. These trap density reductions suggest that HRL-fabricated devices may exhibit improved electronic properties, such as lower charge noise, increased carrier mobility, more stable threshold voltages, and decreased leakage currents. This will benefit any devices that need to be close to the surface and would thus otherwise be exposed to increased defect-induced charge noise.

Our spatial determination of hydrogen's role in trap passivation opens the door to examine other sample processing techniques and investigating how they may influence trap density. For example, halogens such as chlorine are being considered as an alternative resist in hydrogen lithography\cite{halogenResist}. Using a different resist, such as chlorine, may help improve the reliability of atomically-precise doping of the Si(100) surface. However it is unknown if such halogens can passivate traps like hydrogen does. Another silicon processing technique that has shown promise for improving device performance is annealing in deuterium rather than in hydrogen\cite{deuteriumAnnealLyding}. The fm-AFM method used in this work could be used to investigate how such sample processing techniques affect trap density on a local scale.

More broadly, the fm-AFM methodology provides a non-destructive, spatially resolved tool for probing defect landscapes in nanoscale devices. As devices continue to scale down and incorporate quantum functionalities, the ability to directly measure and localize interfacial defects will be essential. These capabilities will be valuable not only for understanding defect dynamics but also for guiding fabrication choices in both the semiconductor and quantum technology industries.\\

\section*{Supplementary Material}
See the supplementary material for details of the experimental setup, analysis of trap densities, and other notes.

\section*{Acknowledgments}
P.G. acknowledges NSERC [grant number 223110], INTRIQ and RQMP for funding. This project was partly financially supported by the Engineering and Physical Sciences Research Council (EPSRC) [grant numbers EP/V027700/1, EP/W000520/1 and EP/Z531236/1] and Innovate UK [grant number UKRI/75574]. N.L.K. was partly supported by the EPSRC Centre for Doctoral Training in Advanced Characterisation of Materials [grant number EP/S023259/1] and Nanolayers Research Computing.

\section*{Author Declarations}

\subsection*{Conflict of Interest}
The authors have no conflicts to disclose.

\subsection*{Author Contributions}
A.J.C. and N.L.K. have contributed equally to this work.

\textbf{A.J.C.}: Data curation, Formal analysis, Investigation, Methodology, Software, Writing - original draft, Writing - review \& editing.
\textbf{N.L.K.}: Formal analysis, Methodology, Resources, Software, Writing - original draft, Writing - review \& editing.
\textbf{P.S.}: Resources.
\textbf{N.S.}: Methodology.
\textbf{W.A.B.}: Supervision, Writing - review \& editing.
\textbf{N.J.C.}: Conceptualization, Funding acquisition, Supervision, Writing - review \& editing.
\textbf{T.J.Z.S.}: Conceptualization, Resources, Supervision, Writing - review \& editing.
\textbf{P.G.}: Conceptualization, Funding acquisition, Supervision, Writing - review \& editing.

\section*{Data Availability}

The data that support the findings of this study are available from the corresponding author upon reasonable request. The code for the machine learning part of this paper (synthetic data generation, training, and testing) can be found at https://github.com/nickkolev97/SLSQ.

\nocite{*}
\providecommand{\noopsort}[1]{}\providecommand{\singleletter}[1]{#1}%

\clearpage

\clearpage
\pagestyle{empty} 

\begin{center}
\textbf{Hydrogen Passivation Effects on Spatially Resolved Charge Trap Densities in Si(100)-SiO$_2$}

Adam J. Czarnecki, Nikola L. Kolev, Patrick See, Nick Sullivan, Wyatt A. Behn, Neil J. Curson,
Taylor J.Z. Stock, and Peter Grütter
\end{center}

\renewcommand{\thefigure}{S\arabic{figure}}
\setcounter{figure}{0}

\renewcommand{\thetable}{S\arabic{table}}
\setcounter{table}{0}

\subsection{Experimental Setup}
All measurements were conducted at room temperature with a UHV ($\sim10^{-10}$ mbar) JEOL JSPM-4500A system. The fm-AFM used a beam deflection detection mechanism, and a Nanosensors platinum-iridium coated silicon cantilever (PPP-NCHPt, $\sim 250$ kHz resonant frequency, $\sim 40$ N/m spring constant, $Q\approx 18000$, and 6 nm oscillation amplitude).

\subsection{Counting Rings with a Neural Network} \label{sup_inf_NN}

\subsubsection{Synthetic Data Generation and Model Training}
Since the experimental images in this work are relatively simple, a synthetic dataset was generated to closely resemble the real data and was used to train a neural network for trap counting. The synthetic image generation pipeline involved the following steps: (1) initialize a black canvas, (2) add a random number of blurred white circles with varying intensity and size, (3) add a random number of blurred white rings representing the traps, (4) incorporate AFM-like noise, and (5) adjust the contrast to match the experimental data. The AFM-like noise included Gaussian noise, scan line artifacts, scan line offsets, and image blurring. This pipeline allows for significant variability by sampling parameters (e.g. ring radius, number of circles) from predefined distributions. Further details on the parameter values and generation functions are available in the project's GitHub repository: https://github.com/nickkolev97/SLSQ. Bounding boxes were used as ground truth labels for training, rather than segmentation masks, because overlapping trap rings made pixel-level separation ambiguous. Segmentation would not have reliably distinguished overlapping structures as distinct objects.

\begin{figure}[!h]
\includegraphics[scale=0.41]{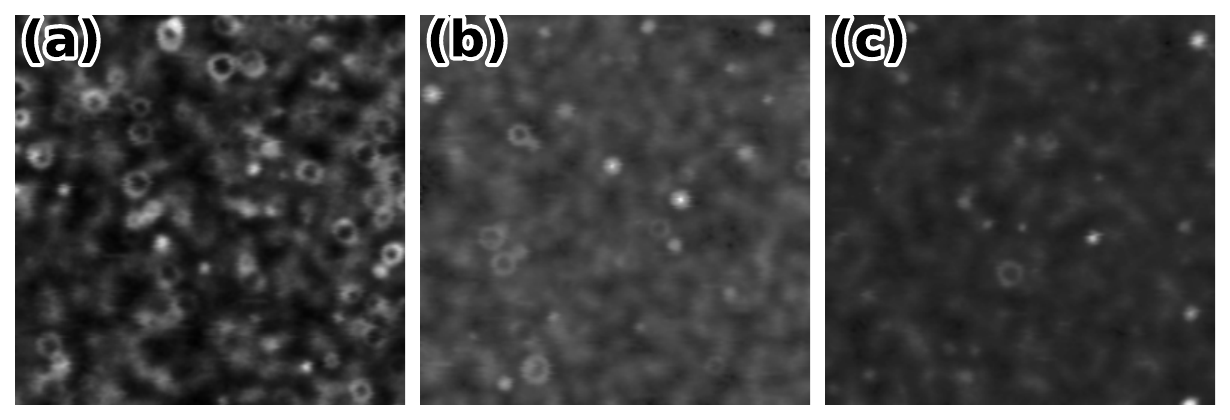} 
\caption{\label{fig:synthetic_examples} Examples of the synthetic data generated for the model training. }
\end{figure}

A total of 4,000 images were generated for training and 1,000 for validation. As the primary goal was generalization to real experimental data, a synthetic test set was not created. The model was trained for 150 epochs and a batch size of 16, with a standard gradient descent (SGD) optimizer that started with a learning rate of 0.01 and a linear learning rate scheduler that decays the learning rate to 0.0001 by the end of the training.

\subsubsection{Model Selection and Evaluation}

The YOLOv8 object detection model developed by Ultralytics \cite{yolov8} was selected to locate and count trap rings. Three variants—nano, medium, and large—were trained, with the medium model showing the best performance on the validation set. YOLOv8 provides four performance metrics: precision (P), recall (R), mean average precision at intersection over union (IoU) of threshold 0.5 (mAP@0.5), and mAP averaged over IoU thresholds from 0.5 to 0.95 (mAP@0.5:0.95).

\textbf{Precision and Recall:}
\begin{equation}
\text{Precision} = \frac{TP}{TP + FP}
\end{equation}

\begin{equation}
\text{Recall} = \frac{TP}{TP + FN}
\end{equation}

\textbf{Mean Average Precision at IoU = 0.5 (mAP@0.5):}
\begin{equation}
\text{mAP@0.5} = \frac{1}{N} \sum_{i=1}^{N} \text{AP}_i(\text{IoU} \geq 0.5)
\end{equation}

\textbf{Mean Average Precision over IoU thresholds from 0.5 to 0.95 (mAP@0.5:0.95):}
\begin{equation}
\text{mAP@0.5:0.95} = \frac{1}{10N} \sum_{j=1}^{10} \sum_{i=1}^{N} \text{AP}_i(\text{IoU} = 0.5 + 0.05 \times (j - 1))
\end{equation}

Where:
\begin{itemize}
    \item $TP$ = True Positives
    \item $FP$ = False Positives
    \item $FN$ = False Negatives
    \item $N$ = Number of object classes
    \item $\text{AP}_i$ = Average Precision for class $i$
\end{itemize}

\begin{figure}[!h]
\includegraphics[scale=0.41]{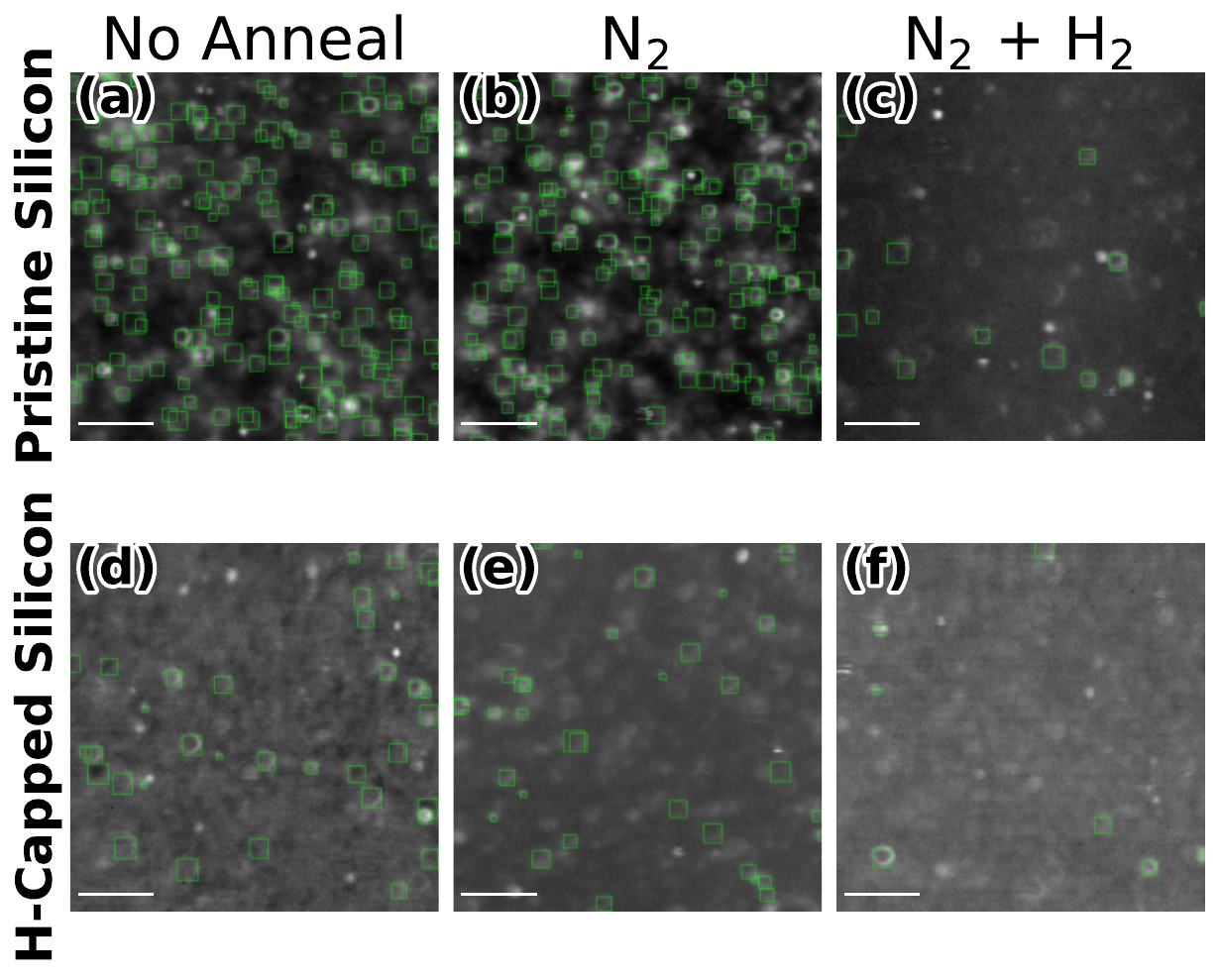} 
\caption{\label{fig:ML-preds} Prediction of medium sized YOLOv8 model on our real data. Not all the AFM images were square, the smallest being 625 nm $\times$ 800 nm, so for an even comparison, and to preserve the aspect ratio, we cropped all images to 625 nm $\times$ 625 nm before inference. Scale bars are 125 nm.}
\end{figure}

These results are summarized in Table \ref{tab:acc}. However, as these metrics reflect performance on synthetic data, they do not directly indicate performance on experimental images. Since a labeled experimental dataset was not created (as this would undermine the motivation for a machine learning-based solution), direct quantitative validation on real data is not available. In Fig. \ref{fig:ML-preds}, we provide the output of the YOLOv8 model on the six images.
\begin{table}[!h]
    \centering
    \begin{tabular}{|c|c|c|c|}
        \hline
        P & R & mAP@0.5 & mAP@0.95 \\
        \hline
        0.85 & 0.66 & 0.76 & 0.56 \\
        \hline
    \end{tabular}
    \caption{Accuracy metrics for YOLOv8 model on the validation set.}
    \label{tab:acc}
\end{table}

\subsubsection{Uncertainty Estimation}
The original YOLOv8 model provided by Ultralytics does not provide an uncertainty with its predictions. To get an uncertainty, we add some randomness to our YOLOv8 model by inserting dropout layers (with probabilities of 0.4) at 6 places in the network (exact locations can be found in the .yaml file in the GitHub repository). By keeping the dropout layers on during inference, the network can be approximated as a Bayesian approximation for deep Gaussian processes, allowing for uncertainty estimation \cite{gal2016dropout}. In practice, the prediction for each image is run $n$ times ($n \in \mathbb{N}$), the mean number of traps found is used as the final prediction, and the standard deviation in this is the error. In this work, $n=20$.

\subsection{Image Binarization using Otsu Thresholding}

\begin{figure}[h]
\includegraphics[scale=0.41]{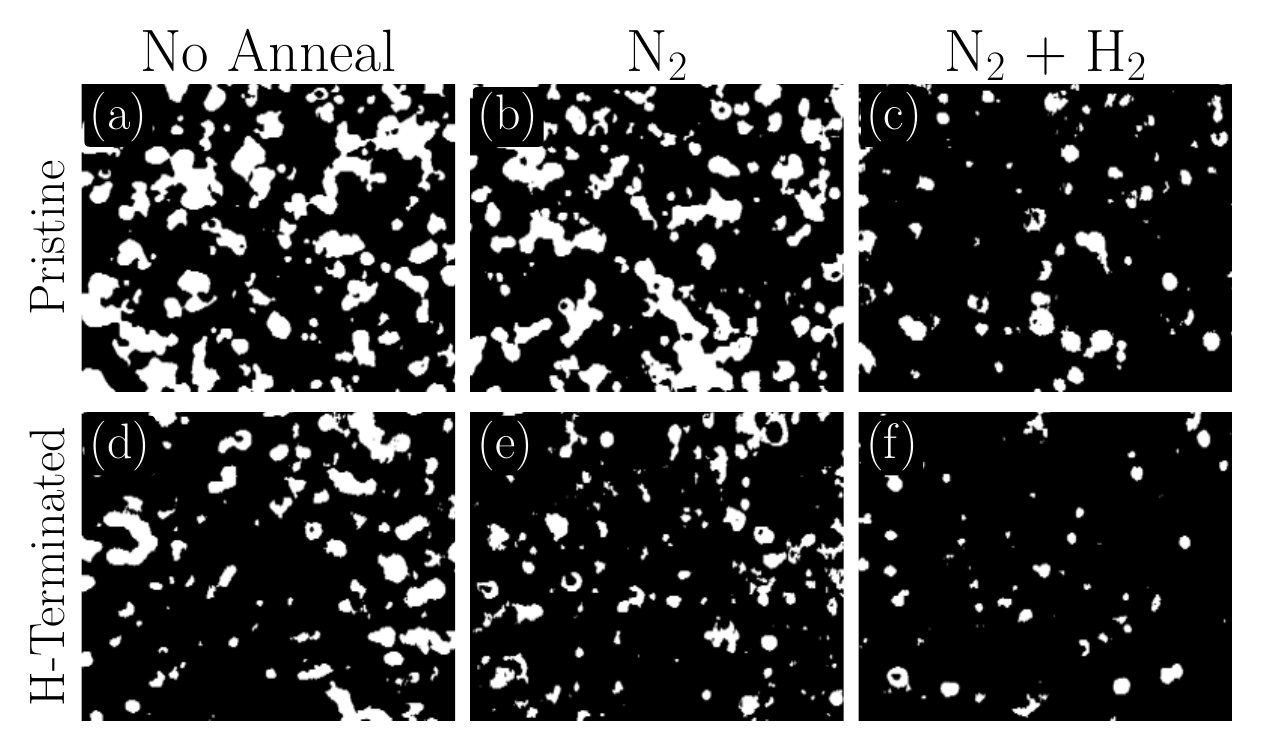}
\caption{\label{fig:binary} Binarization of the dissipation scans shown in Fig.~\ref{fig:pristine-scans} reveal the areas occupied by rings (white areas). Binarization was done using Otsu thresholding.}
\end{figure}

Binarization of the dissipation scans using Otsu thresholding was done to calculate the area occupied by rings. The binary images of the dissipation scans shown in Fig.~\ref{fig:pristine-scans} are shown in Fig.~\ref{fig:binary}.

\subsection{Second Derivative Images}

\begin{figure}[h]
\includegraphics[scale=0.41]{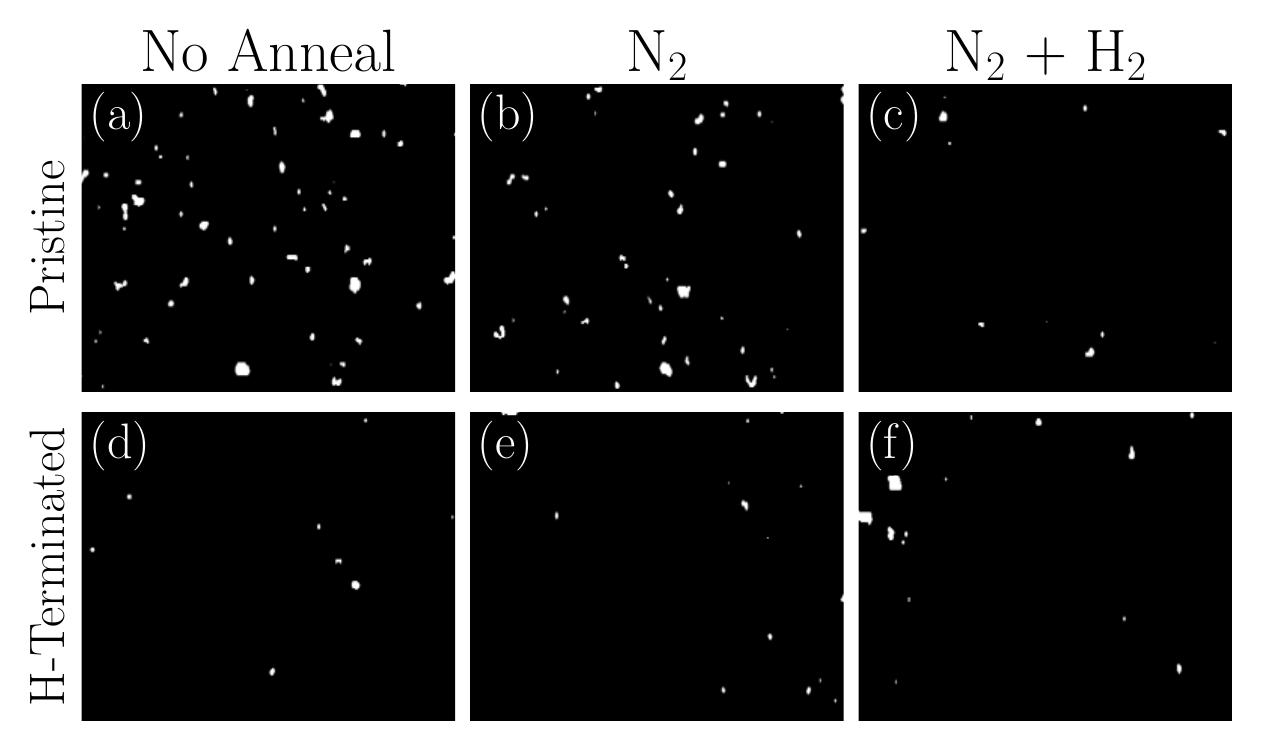}
\caption{\label{fig:2_level_binary} Binarized images of the second derivative taken along the y axis of the dissipation scans in Fig.~\ref{fig:pristine-scans} reveal the locations of two-level traps (white areas).}
\end{figure}

Fig.~\ref{fig:2_level_binary} shows binary images of the second derivative taken along the y axis across all dissipation samples. Binarization was done to make the large second derivative areas more prominent, which correspond to two-level traps. The quantification of two-level traps was done by counting the number of discrete white areas rather than by finding the total area, as there was very little overlap between individual two-level traps, and white area sizes varied greatly.

\subsection{Manual Counting of Rings}
Manual counting of rings was done to verify the evaluation of trap density via area coverage.

\begin{table}[!h]
    \centering
    \begin{tabular}{|c|c|c|c|}
        \hline
         & No Anneal & N$_2$ & N$_{2}$+H$_{2}$\\
        \hline
        Pristine & 250 & 230 & 80 \\
        \hline
        H-Terminated & 120 & 85 & 35 \\
        \hline
    \end{tabular}
    \caption{Approximate number of rings counted manually for each image shown in Fig.~\ref{fig:pristine-scans}.}
    \label{tab:acc}
\end{table}

\clearpage

\end{document}